\documentclass[12pt]{iopart}
\usepackage{epsfig}
\usepackage{graphicx}
\usepackage{cite}

\newcommand{\beq}{\begin{equation}}
\newcommand{\eeq}{\end{equation}}
\newcommand{\bqa}{\begin{eqnarray}}
\newcommand{\eqa}{\end{eqnarray}}

\begin{document}

\title{Thermalization and the chromo-Weibel instability}

\author{Michael Strickland}

\address{Frankfurt Institute for Advanced Studies \\
  Johann Wolfgang Goethe - Universit\"at Frankfurt \\
  Max-von-Laue-Stra\ss{}e~1 \\
  D-60438 Frankfurt am Main, Germany }

\begin{abstract}

Despite the apparent success of ideal hydrodynamics in describing the 
elliptic flow data which have been produced at Brookhaven National 
Lab's Relativistic Heavy Ion Collider, one lingering question remains: 
is the use of ideal hydrodynamics at times $t < 1$ fm/c justified?  In 
order to justify its use a method for rapidly producing isotropic 
thermal matter at RHIC energies is required. One of the chief 
obstacles to early isotropization/thermalization is the rapid 
longitudinal expansion of the matter during the earliest times
after the initial nuclear impact.  As a result of this 
expansion the parton distribution functions become locally anisotropic 
in momentum space. In contrast to locally isotropic plasmas 
anisotropic plasmas have a spectrum of soft unstable modes which are 
characterized by exponential growth of transverse chromo-magnetic/-electric 
fields at short times. This instability is the QCD analogue 
of the Weibel instability of QED. Parametrically the chromo-Weibel 
instability provides the fastest method for generation of soft 
background fields and dominates the short-time dynamics of the system.

\end{abstract}

\section{Introduction}

With the ongoing ultrarelativistic heavy-ion collision experiments at the 
Relativistic Heavy-Ion Collider (RHIC) and planned the Large Hadron Collider 
(LHC) physicists hope to produce and study the properties of a thermalized 
quark-gluon plasma (QGP) which is expected to be formed when the temperature of 
nuclear matter is raised above its critical value, $T_c \sim 200$ MeV $\sim 
10^{12}$ K. Given the small size and short lifetime of the matter created in an 
ultrarelativistic heavy-ion collision this is not trivially accomplished.  One 
of the chief obstacles to thermalization in ultrarelativistic heavy-ion 
collisions is the rapid longitudinal expansion of the matter created in the 
central rapidity region. If the matter expands too quickly then there will not 
be sufficient time for its constituents to interact and thermalize. During the 
first 1 fm/c after the nuclear impact the longitudinal expansion causes the 
created matter to become much colder in the longitudinal direction than in the 
transverse directions~\cite{Baier:2000sb}, corresponding to $\langle p_L^2 
\rangle \ll \langle p_T^2\rangle$ in the local rest frame.  After this initial 
period of longitudinal cooling, the expansion slows and one can then ask what 
are the dominant mechanisms for driving the system towards an isotropic thermal 
QGP.

The evolution of the partonic matter created during a high-energy nuclear 
collision was among the questions which the ``bottom-up'' thermalization 
scenario~\cite{Baier:2000sb} attempted to answer. For the first time, it 
addressed the dynamics of soft modes (fields) with momenta much below $Q_s$ 
coupled to the hard modes (particles) with momenta on the order of $Q_s$ and 
above \cite{Mueller:2002kw,Iancu:2003xm,McLerran:2005kk}. However, it has 
emerged recently that one of the assumptions made in this model was not correct. 
The debate centers around the fact that the 
bottom-up scenario implicitly assumed that the underlying soft gauge modes 
behave the same in an anisotropic plasma as in an isotropic one.  
However, to be self-consistent one must determine the collective modes which are 
relevant for an anisotropic plasma and use those.  In the case of gauge theories 
this turns out to be a qualitative rather than quantitative correction since in 
anisotropic QCD plasmas the most important collective mode corresponds to an 
instability to transverse chromo-magnetic field fluctuations 
\cite{Weibel:1959,Mrowczynski:1993qm,Mrowczynski:1994xv,Romatschke:2003ms}. 
This instability is the QCD analogue of the QED {\em Weibel 
instability}~\cite{Weibel:1959}.  There are now a plethora of analytic and
numerical studies of the {\em chromo-Weibel instability} \cite{Romatschke:2003ms,Arnold:2003rq,Romatschke:2004jh,%
Mrowczynski:2004kv,Romatschke:2006nk,Rebhan:2004ur,Arnold:2005vb,Rebhan:2005re,%
Romatschke:2006wg,Arnold:2005ef,Arnold:2005qs,Dumitru:2005gp,Dumitru:2006pz,Schenke:2006fz,Schenke:2006xu}.

In particular in the last few years there have been significant advances in the understanding of 
non-Abelian soft-field dynamics in anisotropic plasmas within the HL 
framework~\cite{Rebhan:2004ur,Arnold:2005vb,Rebhan:2005re,Romatschke:2006wg}.  
The HL framework is equivalent to the collisionless Vlasov 
theory of eikonalized hard particles, i.e.\ the particle trajectories 
are assumed to be unaffected (up to small-angle scatterings with 
$\theta \sim g$) by the induced background field. It is strictly 
applicable only when there is a large scale separation between the 
soft and hard momentum scales.  Even with these simplifying 
assumptions, HL dynamics for non-Abelian theories is complicated by 
the presence of non-linear gauge-field interactions.  
These non-linear interactions become important when the vector 
potential amplitude is on the order of $A_{\rm non-Abelian} \sim 
p_{\rm s}/g \sim \sqrt{f_h} p_h$, where $p_h$ is the characteristic 
momentum of the hard particles, e.g. $p_h \sim Q_s$ for CGC initial 
conditions, $f_h$ is the angle-averaged occupancy at the hard scale, 
and $p_s$ is the characteristic soft momentum of the fields ($p_s \sim 
g \sqrt{f_h} p_h$).  In QED there are no gauge field self-interactions and the 
fields grow exponentially until $A_{\rm Abelian} \sim p_h/g$ at which 
point the hard particles undergo large-angle deflections by the soft 
background field causing the particles to rapidly isotropize.  In fact, in QED 
the Weibel instability is the fastest process driving plasma 
isotropization.  In QCD, however, the effect of the non-linear gauge 
self-interactions affects the system's dynamics primarily slowing down 
instability-driven particle isotropization.

To include the effects of gauge self-interactions numerical studies of the time 
evolution of the gauge-covariant HL equations of motion are required. Recent 
numerical studies of HL gauge dynamics for SU(2) gauge theory indicate that for 
{\em moderate} anisotropies the gauge field dynamics changes from exponential 
field growth indicative of a conventional Abelian plasma instability to linear 
growth when the vector potential amplitude reaches the non-Abelian scale, 
$A_{\rm non-Abelian} \sim p_{\rm h}$ \cite{Arnold:2005vb,Rebhan:2005re}. This 
linear growth regime is characterized by a turbulent cascade of the energy 
pumped into the soft modes by the instability to higher-momentum plasmon-like 
modes \cite{Arnold:2005ef,Arnold:2005qs}.  These results indicate that there is 
a fundamental difference between Abelian and non-Abelian plasma instabilities in 
the HL limit.

In addition to numerical studies in the HL limit there have been numerical results 
from the solution to the full non-linear Vlasov equations for anisotropic 
plasmas \cite{Dumitru:2005gp,Dumitru:2006pz}.  This approach can be shown to 
reproduce the HL effective action in the weak-field 
approximation~\cite{Kelly:1994ig,Kelly:1994dh,Blaizot:1999xk}; however, when 
solved fully the approach goes beyond the HL approximation since the full 
classical transport theory also reproduces some higher $n$-point vertices of the 
dimensionally reduced effective action for static gluons~\cite{Laine:2001my}. 
Numerical solution of the 3d Vlasov equations show that chromo-instabilities 
persist beyond the HL limit \cite{Dumitru:2006pz}.  Furthermore, 
the soft field spectrum obtained from full Vlasov simulations shows a cascade or 
``avalanche'' of energy deposited in the soft unstable modes in higher momentum 
modes similar to HL dynamics.

\section{Collective Modes of an Anisotropic Quark-Gluon Plasma}

In this section I review the determination of the collective modes of a quark-gluon 
plasma which has a parton distribution function
which is anisotropic in momentum-space.
To simplify things we will assume that $f({\bf p})$ can be obtained 
from an isotropic distribution function by the rescaling of only one direction 
in momentum space.  
In practice this means that, given any isotropic 
parton distribution function $f_{\rm iso}(p)$, we can construct an 
anisotropic version by changing the argument of the isotropic distribution 
function, 
$f({\bf p}) =  f_{\rm iso}\left(\sqrt{{\bf p}^2+\xi({\bf p}\cdot{\bf \hat n})^2}\right)$,
where ${\bf \hat n}$ is the direction of the anisotropy, and $\xi>-1$ is an adjustable 
anisotropy parameter with $\xi=0$ corresponding to the isotropic case. 
Here we will concentrate on $\xi>0$ which corresponds to a 
contraction of the distribution along the ${\bf \hat n}$ direction since this is 
the configuration relevant for heavy-ion collisions at early times, namely
two hot transverse directions and one cold longitudinal direction.
The resulting expression for the gluon polarization tensor is~\cite{Romatschke:2003ms}
\bqa
\Pi^{i j}_{a b}(\omega/k,\theta_n) &=& m_D^2 \, \delta_{a b} \int \frac{d \Omega}{4 \pi} v^{i}%
\frac{v^{l}+\xi({\bf v}\cdot\hat{\bf n}) \hat{n}^{l}}{%
(1+\xi({\bf v}\cdot\hat{\bf n})^2)^2} 
\left( \delta^{j l}+\frac{v^{j} k^{l}}{K\cdot V + i \epsilon}\right) ,
\eqa
where $K=(\omega,{\bf k})$, $V=(1,{\bf p}/p)$,
$\cos\theta_n \equiv \hat{\bf k}\cdot\hat{\bf n}$ and $m_D^2>0$. 
The isotropic Debye mass, $m_D$, depends on 
$f_{\rm iso}$ but is parametrically $m_D \sim g\, p_h$.

\begin{figure}
\flushright\includegraphics[width=8.5cm]{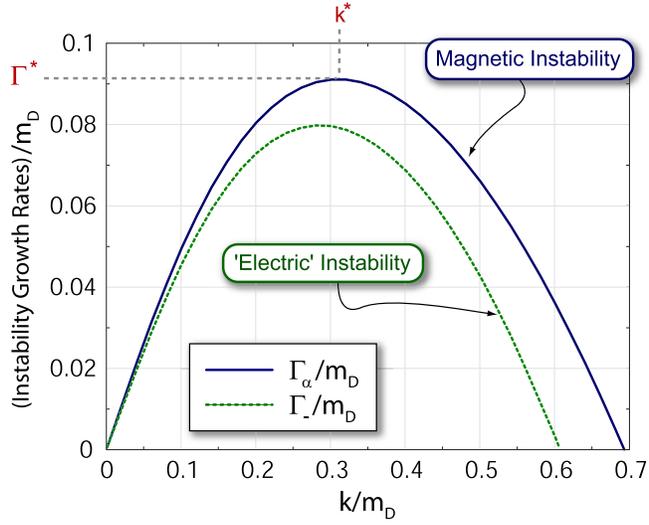}\hspace{1.9cm}
\caption{Instability growth rates as a function of wave number for $\xi=10$ and 
$\theta_n=\pi/8$.  Note that both growth rates vanish at $k=0$ and have a 
maximum $\Gamma^*\sim m_D/10$ at $k^*\sim m_D/3$.}
\label{unstablemodesfig}
\end{figure}

For anisotropic systems ($\xi\neq0$) one can solve for the collective 
modes of the plasma and the result is that there is one additional 
stable mode compared to the isotropic case and, more importantly, that 
there are now purely imaginary solutions in the lower- and upper-halves 
of the complex plane.  These new solutions correspond to damped 
and unstable modes, respectively. We can determine the growth rate for 
these unstable modes by taking $\omega\rightarrow i\Gamma$ and then 
solving the resulting dispersion relations for 
$\Gamma(k)$~\cite{Romatschke:2003ms}. Typical dispersion relations are 
shown in Fig.~\ref{unstablemodesfig}.  As can been seen from this 
figure for $\xi>0$ there are two types of unables modes corresponding 
to magnetic and electric field instabilities.  The magnetic 
instability has a slightly higher growth rate than the electric one 
and so will dominate the dynamics of the system at short times. 
Additionally, since the unstable mode growth rate has a maximum at a wave 
number $k^*$ this means that modes with this wavenumber will be 
predominate.

\section{Numerical Solution of Hard-Loop Dynamics}%

It is possible to go beyond an analysis of gluon polarization tensor 
to a full effective field theory for the soft modes and then solve 
this numerically.  The effective field theory for the soft modes that 
is generated by integrating out the hard plasma modes at one-loop 
order and in the approximation that the amplitudes of the soft gauge 
fields obey $A \ll |\mathbf p|/g$ is that of the gauge-covariant 
collisionless Boltzmann-Vlasov equations \cite{Blaizot:2001nr}. In 
equilibrium, the corresponding (nonlocal) effective action is the so-
called hard-thermal-loop effective action which has a simple 
generalization to plasmas with anisotropic momentum distributions 
\cite{Mrowczynski:2004kv}. For the general non-equilibrium situation 
the resulting equations of motion are
\begin{eqnarray}
D_\nu(A) F^{\nu\mu} &=& -g^2 \int {d^3p\over(2\pi)^3} {1\over2|\mathbf p|} \,p^\mu\, 
						 {\partial f(\mathbf p) \over \partial p^\beta} W^\beta(x;\mathbf v) \, , \nonumber \\
F_{\mu\nu}(A) v^\nu &=& \left[ v \cdot D(A) \right] W_\mu(x;\mathbf v) \, , 
\label{eom}
\end{eqnarray}
where $f$ is a weighted sum of the quark and gluon distribution functions 
and $v^\mu\equiv p^\mu/|\mathbf p|=(1,\mathbf v)$.

These equations include all hard-loop resummed propagators and 
vertices and are implicitly gauge covariant.  At the expense of 
introducing a continuous set of auxiliary fields $W_\beta(x;\mathbf 
v)$ the effective field equations are also local.  These equations of 
motion are then discretized in space-time and ${\mathbf v}$, and 
solved numerically.  The discretization in ${\mathbf v}$-space 
corresponds to including only a finite set of the auxiliary fields 
$W_\beta(x;\mathbf v_i)$ with $1 \leq i \leq N_W$. For details on the 
precise discretizations used see 
Refs.~\cite{Arnold:2005vb,Rebhan:2005re}.

\begin{figure}[t]
\flushright\includegraphics[width=9cm]{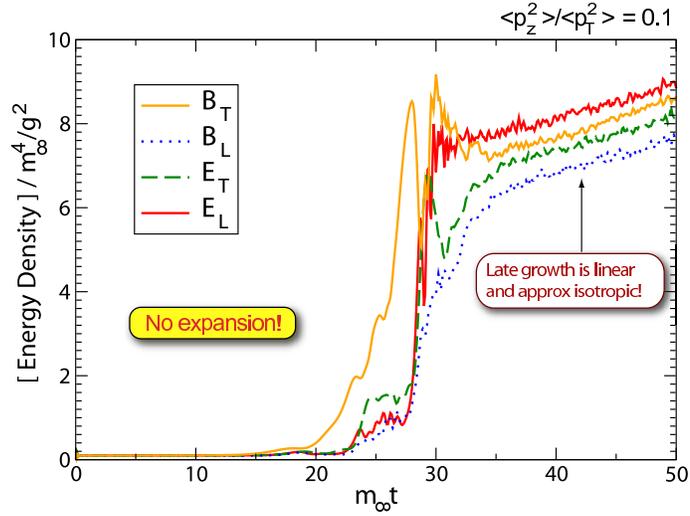}\hspace{1.9cm}
\caption{
Plot of typical energy densities observed in non-expanding three-dimensional 
hard-loop simulation of the soft-fields generated in an anisotropic plasma with 
$\xi=10$.  Shows transition from exponential growth with preference for 
transverse magnetic fields to linear isotropic growth.
\label{3dHLfig}}
\end{figure}

\subsection{Discussion of Numerical Hard-Loop Simulation Results}

During the process of instability growth the soft gauge fields get the 
energy for their growth from the hard particles and, of course, the 
total energy is conserved.  In an Abelian plasma the energy deposited 
in soft fields grows exponentially until the energy in the soft fields 
is of the same order as the energy remaining in the hard particles at 
which point the back-reaction of the fields on the particle motion 
causes rapid isotropization.  As mentioned above in a non-Abelian 
plasma the situation is quite different and one must rely on numerical 
simulations due to the presence of strong gauge field self-interactions. 

\begin{figure}[t]
\flushright\includegraphics[width=10.5cm]{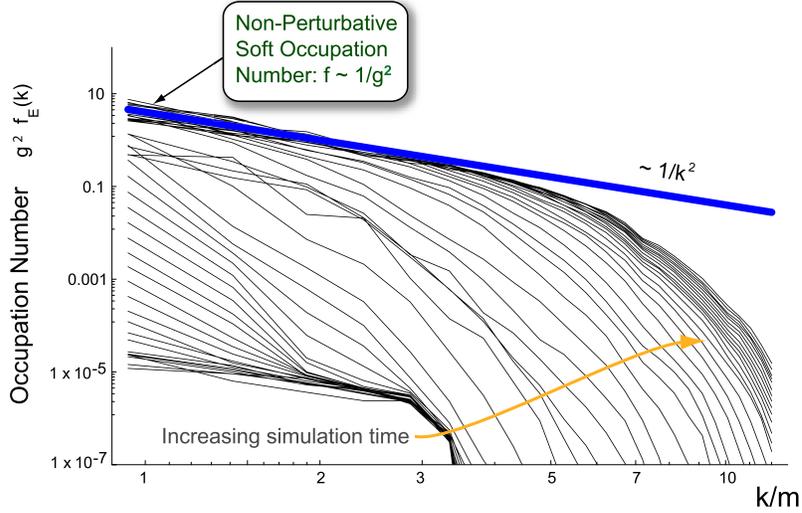}\hspace{1.9cm}
\caption{
Field mode spectrum for an SU(2) run showing saturation of soft field growth at $f \sim 1/g^2$
and an associated cascade of energy to the UV as the simulation time increases.
\label{cascadefig}}
\end{figure}

In Fig.~\ref{3dHLfig} I have plotted the time dependence of the 
chromo-magnetic/-electric energy densities obtained from a 3+1 
dimensional from a typical HL simulation run initialized with ``weak'' 
random color noise with $\xi=10$ ~\cite{Rebhan:2005re}.  As can be 
seen from this figure at $m_\infty\,t \simeq 30$ there is a change 
from exponential to linear growth. Another interesting feature of the 
isotropic linear growth phase is that it exhibits a cascade of energy 
pumped into the unstable soft modes to higher energy plasmon like 
modes.  This is demonstrated in Fig.~\ref{cascadefig} which shows the 
soft gauge field spectrum as a function of momentum at different 
simulation times along with the estimated scaling coefficient of the 
spectrum~\cite{Arnold:2005ef}. From Fig.~\ref{3dHLfig} we can conclude 
that the chromo-Weibel instability will be less efficient at 
isotropizing a QCD plasma than the analogous Weibel instability seen 
in Abelian plasmas due to the slower than exponential growth at late 
times. On the positive side, from a theoretical perspective 
``saturation'' at the soft scale implies that one can still apply the 
hard-loop effective theory self-consistently to understand the 
behavior of the system in the linear growth phase.
 
One caveat is that the latest published HL simulation results 
\cite{Arnold:2005vb,Rebhan:2005re} are for distributions with a finite 
${\mathcal O}(1\!\!\rightarrow\!\!10)$ anisotropy due to computational 
limitations and the observed saturation seems to imply that for weak 
anisotropies field instabilities will not rapidly isotropize the hard 
particles.  This means, however, that due to the continued expansion 
of the system the anisotropy will increase.  It is therefore necessary 
to study the hard-loop dynamics in an expanding system.  Naively, one 
expects this to change the growth from $\exp(\tau)$ to 
$\exp(\sqrt\tau)$ at short times but there is no clear expectation of 
what will happen in the linear regime.  A significant advance in this 
regard has occurred recently for a U(1) gauge theory 
\cite{Romatschke:2006wg}.  Work is underway to do the same for non-Abelian 
gauge theories.

\section{Beyond Hard-Loops}
\label{sec_WYMeqs}

It is also to possible to go beyond the hard-loop approximation and solve
instead the full classical transport equations in three dimensions \cite{Dumitru:2006pz}.
The Vlasov transport
equation for hard gluons with non-Abelian color charge $q^a$ in the
collisionless approximation are~\cite{Wong:1970fu,Heinz:1983nx},
\begin{equation}
 p^{\mu}[\partial_\mu - gq^aF^a_{\mu\nu}\partial^\nu_p
    - gf_{abc}A^b_\mu q^c\partial_{q^a}]f(x,p,q)=0 \, ,   \label{Vlasov}
\end{equation}
were $f(t,\bf{x},\bf{p},q^a)$ denotes the single-particle phase space
distribution function.

The Vlasov equation is coupled self-consistently to the Yang-Mills
equation for the soft gluon fields,
\begin{equation}
 D_\mu F^{\mu\nu} = J^\nu = g \int \frac{d^3p}{(2\pi)^3} dq \,q\,
 v^\nu f(t,\bf{x},\bf{p},q)~, \label{YM}
\end{equation}
where again $v^\mu\equiv(1,{\bf p}/p)$. These equations reproduce the
hard-loop effective action near
equilibrium~\cite{Kelly:1994ig,Kelly:1994dh,Blaizot:1999xk}. However, 
the full classical transport
theory~(\ref{Vlasov},\ref{YM}) also reproduces some higher $n$-point
vertices of the dimensionally reduced effective action for
static gluons~\cite{Laine:2001my} beyond the hard-loop approximation. The
back-reaction of the long-wavelength fields on the hard particles
(``bending'' of their trajectories) is, of course, taken into account,
which is important for understanding particle dynamics in strong
fields.  For details of the 
numerical implementation used see Ref.~\cite{Dumitru:2006pz}.

\begin{figure}[t]
\vspace{3mm}
\flushright\includegraphics[width=11.2cm]{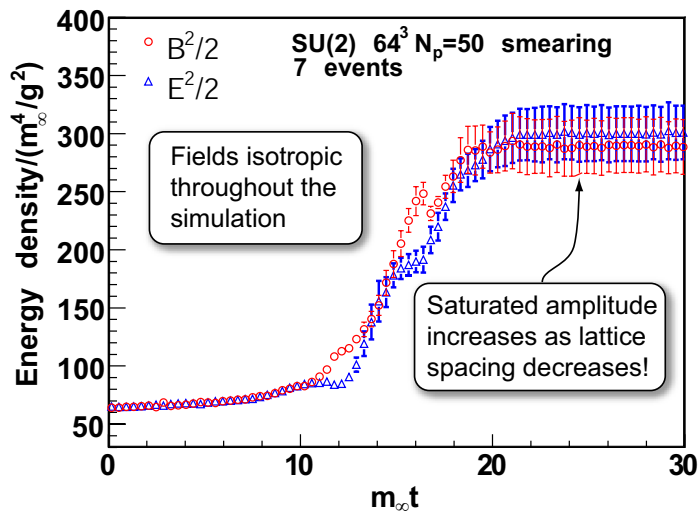}
\vspace{-3mm}
\caption{
Time evolution of the field energy densities for $SU(2)$
gauge group resulting from a highly anisotropic initial particle
momentum distribution.  Simulation parameters are $L=5$~fm, $p_h=16$~GeV, $g^2\,n_g=10/$fm$^3$, $m_\infty=0.1$~GeV.}
\label{cpicfig}
\end{figure}

In Fig.~\ref{cpicfig} I present the results of a three-dimensional Wong-Yang-Mills 
(WYM) simulation published in Ref.~\cite{Dumitru:2006pz}.  The figure 
shows the time evolution of the field energy densities for $SU(2)$ gauge group 
resulting from a highly anisotropic initial particle momentum distribution.
The behavior shown in Fig.~\ref{cpicfig} indicates that the results obtained 
from the hard-loop simulations and direct numerical solution of the WYM 
equations are qualitatively similar in that both show that for non-Abelian 
gauge theories there is a saturation of the energy transferred to the soft modes 
by the gauge instability.  Although I don't show it here the corresponding 
Coulomb gauge fixed field spectra show that the field
saturation is accompanied by an ``avalanche'' of energy transferred to soft 
field modes to higher frequency field modes with saturation occurring when the 
hardest lattice modes are filled ~\cite{Dumitru:2006pz}.
A thorough analytic understanding of this ultraviolet avalanche is lacking 
at this point in time although some advances in this regard have been made 
recently \cite{Mueller:2006up}.

\section{Outlook}

An important open question is whether quark-gluon plasma instabilities and/or 
the physics of anisotropic plasmas in general play an important phenomenological 
role at RHIC or LHC energies.  In this regard the recent papers of 
Refs.~\cite{Romatschke:2003vc,Romatschke:2004au,Schenke:2006fz,Romatschke:2006bb} 
provide theoretical frameworks which can be used to calculate the impact of 
anisotropic momentum-space distributions on observables such as jet shapes and 
the rapidity dependence of medium-produced photons. A concrete example of work 
in this direction is the recent calculation of photon production from an 
anisotropic QGP \cite{Schenke:2006yp}.  The results of that work suggest that it 
may be able possible to determine the time-dependent anisotropy of a QGP by 
measuring the rapidity dependence of high-energy medium photon production.

\section*{References}
\bibliographystyle{iopart-num}
\bibliography{strickland}

\end{document}